# Quantum phase transition and entanglement in Li atom system

SI LiMing[1†] & HOU JiXuan[2]

[1] Department of Electronic Engineering, School of Information Science and Technology, Beijing Institute of Technology, Beijing 100081, China;
[2] Laboratoire de Chimie, UMR 5182 CNRS, Ecole Normale Supérieure de Lyon, 46, Allée d'Italie, F-69364 Lyon Cedex 07, France

**By use of the exact diagonalization method, the quantum phase transition and entanglement in a 6-Li atom system are studied. It is found that entanglement appears before the quantum phase transition and disappears after it in this exactly solvable quantum system. The present results show that the von Neumann entropy, as a measure of entanglement, may reveal the quantum phase transition in this model.**

The quantum phase transition (QPT) driven by a varying magnetic field or pressure is a phase transition of quantum systems[1]. The novel phenomena, which have been found in QPT and do not exist in the thermal phase transition, have stimulated much theoretical investigation on QPT in quantum multi-body systems. Entanglement, as one of the most interesting features of quantum theory, is regarded as a valuable resource in the science of quantum communication[2−8]. Therefore, its measurement is important for quantum communication. The von Neumann entropy, one of the most important measurements, has been used for quantum entanglement[9,10]. Recently, QPT was studied in connection with the entanglement[11−20]. It was proposed that QPT is usually considered to be connected with the point of extremum, singularity or discontinuity for the entanglement. However, a large number of results have been achieved based on the observation that QPT is characterized through nonanalyticities of the density matrix of an appropriate subsystem[21−26].

Because of the complexity of quantum multi-body systems, much attention has been paid to QPT in some small quantum systems (such as a pair of qubits, two-spin system, and three-spin system)[27−30] which are considered as the keys to access to the multi-body QPT. In this paper, the QPT and entanglement (von Neumann entropy) are accurately calculated and the relationship between them is discussed in a 6-Li atom system which is a typical small quantum system and may be easily realized under the experimental condition.

## 1  Quantum phase transition

This part introduces the materials, method and experimental procedure of the author's work, so as to allow others to repeat the work published based on this clear description. Actually, in a 6-Li atom system, the electrons have total spin-1/2 and the nucleus has spin-1. We consider the system

†Corresponding author (email: siliming100@yahoo.com.cn)

spin-1/2 and spin-1 in a magnetic field with the corresponding Hamiltonian:

$$H = J(\mathbf{s}_1 \cdot \mathbf{s}_2) - B(s_{1x} + s_{2x}) = J(s_x^1 s_x^2 + s_y^1 s_y^2 + s_z^1 s_z^2) - B(s_x^1 + s_x^2), \tag{1}$$

where $\mathbf{s}_1$ refers to the Pauli matrices for spin-1/2 and $\mathbf{s}_2$ denotes the spin operators for spin-1,

$$s_x^1 = \begin{bmatrix} 0 & 1 \\ 1 & 0 \end{bmatrix}, \quad s_y^1 = \begin{bmatrix} 0 & -i \\ i & 0 \end{bmatrix}, \quad s_z^1 = \begin{bmatrix} 1 & 0 \\ 0 & -1 \end{bmatrix},$$

$$s_x^2 = \frac{1}{\sqrt{2}} \begin{bmatrix} 0 & 1 & 0 \\ 1 & 0 & 1 \\ 0 & 1 & 0 \end{bmatrix}, \quad s_y^2 = \frac{1}{\sqrt{2}} \begin{bmatrix} 0 & -i & 0 \\ i & 0 & -i \\ 0 & i & 0 \end{bmatrix}, \quad s_z^2 = \begin{bmatrix} 1 & 0 & 0 \\ 0 & 0 & 0 \\ 0 & 0 & -1 \end{bmatrix},$$

and $B$ is the strength of magnetic field. In this paper, the spin-1/2 subsystem is denoted by $|\uparrow\rangle$ and $|\downarrow\rangle$ while the spin-1 subsystem is denoted by $|\Uparrow\rangle$, $|0\rangle$ and $|\Downarrow\rangle$, then the whole system may be expressed by six base vectors: $|\uparrow\Uparrow\rangle$, $|\uparrow 0\rangle$, $|\uparrow\Downarrow\rangle$, $|\downarrow\Uparrow\rangle$, $|\downarrow 0\rangle$, and $|\downarrow\Downarrow\rangle$.

The matrix form of the Hamiltonian is written as

$$H = \begin{bmatrix} J & -\dfrac{B}{\sqrt{2}} & 0 & -B & 0 & 0 \\ -\dfrac{B}{\sqrt{2}} & 0 & -\dfrac{B}{\sqrt{2}} & \sqrt{2}J & -B & 0 \\ 0 & -\dfrac{B}{\sqrt{2}} & -J & 0 & \sqrt{2}J & -B \\ -B & \sqrt{2}J & 0 & -J & -\dfrac{B}{\sqrt{2}} & 0 \\ 0 & -B & \sqrt{2}J & -\dfrac{B}{\sqrt{2}} & 0 & -\dfrac{B}{\sqrt{2}} \\ 0 & 0 & -B & 0 & -\dfrac{B}{\sqrt{2}} & J \end{bmatrix}. \tag{2}$$

It is easy to obtain the eigenvectors and eigenvalues of the system, and the eigenvalues are shown in Figure 1. It will be seen that there is a level crossing where an excited level becomes the ground state at $B = 2J$, creating a point of non-analyticity of the ground state energy. And this is the QPT in our 6-Li atom system.

In details, when $B > 2J$, the minimal eigenvalue of this system is $-2B + J$.

The correspondence eigenvector

$$|Y_{B>2J}\rangle = \frac{1}{\sqrt{8}}\left[|\uparrow\Uparrow\rangle + \sqrt{2}|\uparrow 0\rangle + |\uparrow\Downarrow\rangle + |\downarrow\Uparrow\rangle + \sqrt{2}|\downarrow 0\rangle + |\downarrow\Downarrow\rangle\right], \tag{3}$$

and it may be rewritten into the form:

$$|Y_{B>2J}\rangle = \frac{1}{\sqrt{2}}\left[|\uparrow\rangle + |\downarrow\rangle\right] \cdot \frac{1}{2}\left[|\Uparrow\rangle + \sqrt{2}|0\rangle + |\Downarrow\rangle\right], \tag{4}$$

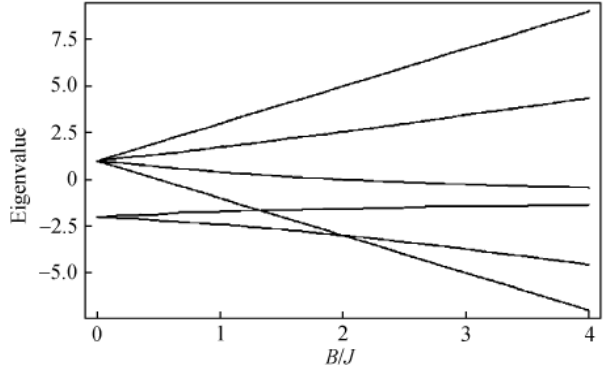

**Figure 1** The eigenvalue of the 6-Li atom system changes with $B/J$. $B$ is the strength of the magnetic field.

Obviously, it is a separable state and the von Neumann entropy is zero.

When $B < 2J$, the minimal eigenvalue of this system is

$$-\frac{1}{2}B - \frac{1}{2}J - \frac{1}{2}\sqrt{B^2 - 2JB + 9J^2},$$

and the correspondence eigenvector is

$$\begin{aligned}|Y_{B<2J}\rangle &= \frac{J^2}{9J^2 + B^2 - (J-B)\sqrt{B^2 - 2JB + 9J^2} - 2JB}\Big[-|\uparrow\Uparrow\rangle \\ &+ \frac{\sqrt{2}(3J - B - \sqrt{B^2 - 2JB + 9J^2})}{2B}|\uparrow 0\rangle + \frac{3J + \sqrt{B^2 - 2JB + 9J^2}}{B}|\uparrow\Downarrow\rangle \\ &- \frac{3J + \sqrt{B^2 - 2JB + 9J^2}}{B}|\downarrow\Uparrow\rangle - \frac{\sqrt{2}(3J - B - \sqrt{B^2 - 2JB + 9J^2})}{2B}|\downarrow 0\rangle + |\downarrow\Downarrow\rangle\Big] \\ &\equiv a_1|\uparrow\Uparrow\rangle + a_2|\uparrow 0\rangle + a_3|\uparrow\Downarrow\rangle + a_4|\downarrow\Uparrow\rangle + a_5|\downarrow 0\rangle + a_6|\downarrow\Downarrow\rangle.\end{aligned} \quad (5)$$

## 2  Entanglement and von Neumann entropy

It is generally believed that there is a kind of connection between the quantum entanglement and QPT. We show that, for this quantum system, when the strength of magnetic field is strong enough ($B > 2J$) the entanglement disappears after the point of QPT; while $B > 2J$ the entanglement exists. Then we will use von Neumann entropy to measure the degree of entanglement. Considering the reduced-density matrix of a subsystem $r_A = \mathrm{tr}_B r = \mathrm{tr}_B |Y\rangle\langle Y|$, the von Neumann entropy $E_n(A)$ may be written as

$$E_n(A) = -\mathrm{tr}[r_A \log_2(r_A)]. \quad (6)$$

When $B < 2J$, the reduced-density matrix is

$$\begin{aligned}r_A = \mathrm{tr}_B r &= \langle\Uparrow|r|\Uparrow\rangle + \langle 0|r|0\rangle + \langle\Downarrow|r|\Downarrow\rangle \\ &= (a_1^2 + a_2^2 + a_3^2)|\uparrow\rangle\langle\uparrow| + (a_4^2 + a_5^2 + a_6^2)|\downarrow\rangle\langle\downarrow|\end{aligned}$$

$$+ (a_1a_4 + a_2a_5 + a_3a_6)|\uparrow\rangle\langle\downarrow| + (a_1a_4 + a_2a_5 + a_3a_6)|\downarrow\rangle\langle\uparrow|$$

$$= \begin{bmatrix} a_1^2 + a_2^2 + a_3^2 & a_1a_4 + a_2a_5 + a_3a_6 \\ a_1a_4 + a_2a_5 + a_3a_6 & a_4^2 + a_5^2 + a_6^2 \end{bmatrix}. \tag{7}$$

By use of eqs. (6) and (7), the corresponding von Neumann entropy and its change with parameter $B/J$ are exactly calculated, and shown in Figure 2. It will be found that when $B < 2J$, the change of von Neumann entropy is not monotonic with the increase of the magnetic field $B$, but it has the maximum von Neumann entropy at $B = J$. When magnetic field $B$ increases to $B > 2J$ or exceeds the point of QPT, entanglement disappears. This clearly testifies the relationship between QPT and von Neumann entropy.

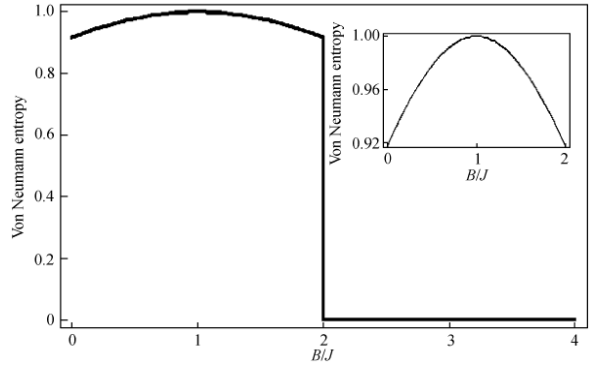

**Figure 2** Entanglement (von Neumann entropy) of the 6-Li atom system changes with $B/J$. $B$ is the strength of the magnetic field.

## 3  Conclusions

A novel method to judge the point of QPT via entanglement is proposed in this paper. QPT in the magnetic field and entanglement by use of von Neumann entropy in a 6-Li atom system are accurately calculated. It is shown that QPT may be well described by the von Neumann entropy in this system, and the point at which entanglement disappears is the point of QPT. In other words, it is interesting that the abruptly disappearing point of entanglement is the consequence of QPT in this system. Therefore, the result may give an effective method for the judgement of QPT or entanglement through another one of them in the quantum system.

1   Sachdev S. Quantum Phase Transitions. Cambridge: Cambridge University Press, 1999
2   Einstein A, Podolsky B, Rosen N. Can quantum-mechanical description of physical reality be considered complete? Phys Rev, 1935, 47: 777－780
3   Bennett C H, Divincenzo D P. Quantum information and computation. Nature, 2000, 404: 247－255
4   Ye M Y, Zhang Y S, Guo G C. Quantum entanglement and quantum operation. Sci China Ser G-Phys Mech Astron, 2008, 51(1): 14－21
5   Ding S C, Jin Z. Review on the study of entanglement in quantum computation speedup. Chin Sci Bull, 2007, 52(16): 2161－2166
6   Di Y M, Liu L. Entanglement capacity of two-qubit unitary operator for rank two mixed states. Sci China Ser G-Phys Mech Astron, 2007, 50(6): 691－697 7   Liu D, Zhang Y, Long G L. Influence of magnetic current on the ground state entanglement in an isotropic transverse XY chain. Prog Nat Sci, 2007, 17: 1147−1151

8   Hu M L, Tian D P. Effects of impurity on the entanglement of the three-qubit Heisenberg XXX spin chain. Sci China Ser G-Phys Mech Astron, 2007, 50(2): 208−214

9   Vedral V, Plenio M B, Rippin M A, et al. Quantifying entanglement. Phys Rev Lett, 1997, 78: 2275−2279

10  Cao W C, Liu D, Pan F, et al. Entropy product measure for multipartite pure states. Sci China Ser G-Phys Mech Astron, 2006, 49(5): 606−615

11  Chen Y, Zanardi P, Wang Z D, et al.. Sublattice entanglement and quantum phase transitions in antiferromagnetic spin chains. New J Phys, 2006, 8: 97

12  Gu S J, Tian G S, Lin H Q. Local entanglement and quantum phase transition in spin models. New J Phys, 2006, 8: 61

13  Wu L A, Sarandy M S, Lidar D A. Quantum phase transitions and bipartite entanglement. Phys Rev Lett, 2004, 93: 250404

14  Latorre J I, Rico E, Vidal G. Ground state entanglement in quantum spin chains. Quant Inf Comput, 2004, 4: 48

15  Yang M F. Reexamination of entanglement and the quantum phase transition. Phys Rev A, 2005, 71: 030302

16  Zhang Y, Cao W C, Long G L. Creation of multipartite entanglement and entanglement transfer via heisenberg interaction. Chin Phys Lett, 2005, 22: 2143−2146

17  Liu D, Zhang Y, Liu Y, et al. Entanglement in the ground state of an isotropic three-qubit transverse XY chain with energy current. Chin Phys Lett, 2007, 24: 8−10

18  Zhang Y, Liu D, Long G L. Ground-state entanglement in a three-spin transverse Ising model with energy current. Chin Phys, 2007, 16: 324−328

19  Zhang Y, Long G L. Ground-state and thermal entanglement in three-spin Heisenberg-XXZ chain with three-spin interaction. Commun Theor Phys, 2007, 48: 249−254

20  Zhang Y, Long G L, Wu Y C, et al. Partial teleportation of entanglement through natural thermal entanglement in two-qubit Heisenberg XXX chain. Commun Theor Phys, 2007, 47: 787−790

21  Wu L A, Sarandy M S, Lidar D A, et al. Linking entanglement and quantum phase transitions via density functional theory. Phys Rev A, 2006, 74: 052335

22  Roscilde T, Verrucchi P, Fubini A, et al. Studing quantum spin systems through entanglement estimators. Phys Rev Lett, 2004, 93: 167203

23  Amico L, Baroni F, Fubini A, Patane D, Tognetti V, Verrucchi P. Divergence of the entanglement range in low-dimensional quantum systems. Phys Rev A, 2006, 74: 022322

24  Venuti C L, Boschi D E C, Roncaglia M, et al. Local measures of entanglement and critical exponents at quantum phase transitions. Phys Rev A, 2006, 73: 010303

25  Zanardi P. Quantum entanglement in fermionic lattices. Phys Rev A, 2002, 65: 042101

26  Anfossi A, Giorda P, Montorsi A. Entanglement in extended hubbard models and quantum phase transitions. Phys Rev B, 2007, 75: 165106

27  Facchi P, Marzolino U, Parisi G, et al. Phase transitions of bipartite entanglement. arXiv: 0712.0015v1 [quant-ph] 30 Nov 2007

28  Pižorn I, Prosen T, Mossmann S, et al. The two-body random spin ensemble and a new type of quantum phase transition. arXiv: 0711.1218v1 [quant-ph] 8 Nov 2007

29  Alvarez G A, Danieli E P, Levstein P R, et al. Environmentally induced quantum dynamical phase transition in the spin swapping operation. J Chem Phys, 2006, 124: 194507

30  Alvarez G A, Levstein P R, Pastawski H M. Signatures of a quantum dynamical phase transition in a three-spin system in presence of a spin envernoment. Phys B, 2007, 398: 438−441